\documentclass[twocolumn]{aastex61}
\usepackage{amsmath}
\submitjournal{ApJ}

\shorttitle{magnetized hot accretion flows}
\shortauthors{Zeraatgari et al.}

\begin{document}

\title{The effects of toroidal magnetic field on the vertical structure of hot accretion flows}

\correspondingauthor{Amin Mosallanezhadi}
\email{fzeraatgari@ustc.edu.cn}
\email{mosallanezhad@ustc.edu.cn}
\email{abbassi@um.ac.ir}

\author{Fatemeh Zahra Zeraatgari}
\affil{Key Laboratory for Research in Galaxies and Cosmology, Department of Astronomy, University of Science and Technology of China, Hefei, Anhui 230026, China}

\author{Amin Mosallanezhad}
\affiliation{Key Laboratory for Research in Galaxies and Cosmology, Department of Astronomy, University of Science and Technology of China, Hefei, Anhui 230026, China}

\author{Shahram Abbassi}
\affiliation{Department of Physics, School of Sciences, Ferdowsi University of Mashhad, 91775-1436 Mashhad, Iran}
\affiliation{School of Astronomy, Institute for Studies in Theoretical Physics and Mathematics, 19395-5531 Tehran, Iran}

\author{Ye-Fei Yuan}
\affiliation{Key Laboratory for Research in Galaxies and Cosmology, Department of Astronomy, University of Science and Technology of China, Hefei, Anhui 230026, China}



\begin{abstract}
We solved the set of two-dimensional magnetohydrodynamic (MHD) equations for optically thin black hole accretion flows incorporating toroidal component of magnetic field. Following global and local MHD simulations of black hole accretion disks, the magnetic field inside the disk is decomposed into a large scale field and a fluctuating field. The effects of the fluctuating magnetic field in transferring the angular momentum and dissipating the energy are described through the usual $ \alpha $ description. We solved the MHD equations by assuming steady state and radially self-similar approximation in $ r-\theta $ plane of spherical coordinate system. 
We found that as the amount of magnetic field at the equatorial plane increases, the heating by the viscosity decreases.
In addition, the maximum amount of the heating by the viscous dissipation is produced at the mid-plane of the disk, while that of the heating by the magnetic field dissipation 
is produced at the surface of the disk.
Our main conclusion is that in terms of the no-outflow solution, thermal equilibrium still exists 
for the strong magnetic filed at the equatorial plane of the disk.

\end{abstract}




\section{Introduction} \label{sec:intro}
It is well known that mass accretion onto a black hole is a common 
process in the universe and is the power source of many active 
phenomena including X-ray binaries (XRBs), active galactic nuclei 
(AGNs) and gamma-ray bursts (GRBs). Based on the temperature 
of accretion flow, two distinct classes of black hole accretion solutions 
exist, i.e., cold and hot flows.

The standard thin disk, commonly named as an $ \alpha$-disk, 
is categorized in cold accretion flows (Shakura \& Sunyaev 1973;
Lynden-Bell \& Pringle 1974; Novikov \& Thorne 1973). In this 
model the disk is considered to be in the limit of a optically thick, 
geometrically thin ($ H/R \ll 1$, where $ H $ is the half thickness 
of the disk in the cylindrical radius $ R $), Keplerian rotating, and 
radially subsonic disk. Moreover, in $ \alpha$-disk typically global 
heat transport is neglected and the energy released via viscosity 
is radiated away locally. Therefore, the accreting flow becomes 
cool very efficiently and cannot produce a high-energy spectrum. 
Additionally, the mass accretion rate of the standard disk is mildly 
low, i.e., $ \dot{M} \lesssim \dot{M}_{\text{crit}} $, where $ \dot{M}_{\text{crit}} $ 
is the critical mass accretion rate. The high/soft state of black hole 
binaries (BHBs) as well as usual luminous AGNs belong to this 
model (see reviews by Pringle 1981; Frank et al. 2002; Kato et al. 2008; 
Abramowicz \& Fragile 2013; Blaes 2014; Lasota 2015 for more details).

When the mass accretion rate becomes extremely high, i.e.,
$ \dot{M} \ga \dot{M}_{\text{crit}} $, the accreting flow turns 
to be optically thick and the energy released cannot radiate 
away locally. Consequently, the radiation is trapped and 
advected with the accreting matter inwardly. This flow with such a high mass 
accretion rate is called slim disk (see, Abramowicz et al. 1988; 
see also Abramowicz et al. 1986; Kato et al. 2008; 
Chen \& Taam 1993; Narayan \& Popham 1993; Katz 1977; 
Begelman 1979; Begelman \& Meier 1982; Eggum et al. 1988 
for more details). The slim disk also belongs to the cold class 
and like standard disk emits blackbody-like radiation. The horizontal 
pressure gradients, radial velocity, and also advective heat 
transport of slim disks are not negligible. This class of 
solutions is applied to the systems such as ultraluminous X-ray 
sources (ULXs), ultraluminous supersoft X-ray sources 
(ULSs), luminous quasars, and narrow-line Seyfert 1 galaxies 
(Fukue 2004; Kato et al. 2008).

In contrast to the cold disk model, hot accretion solution also 
exists when the mass accretion rate is very low ($ \dot{M} \lesssim 
\alpha^{2} \dot{M}_{\text{Edd}} $, where $ \alpha $ is viscous
parameter and $ \dot{M}_{\text{Edd}} $ is the Eddington accretion 
rate). Early work on hot accretion flows was first initiated by Ichimaru (1977)
and Rees et al. (1982), then rediscovered by Narayan \& Yi (1994, 
1995a, 1995b), Abramowicz et et al. 1995. In this model, the disk 
is optically thin and since the cooling mechanism is inefficient, 
therefore the gas temperature becomes extremely high (nearly 
virial). Consequently, the gas pressure acts to puff up the 
inner region of the accretion disk and the disk becomes geometrically 
thick, i.e., $ H/R  \sim 1$. The dynamics and radiative properties 
of hot accretion solution have been investigated in detail and 
this model has widespread applications in various sources, 
such as the supermassive black hole in our Galactic Center, 
Sagittarius A$^{*}$ (Sgr A$^*$), low-luminosity AGNs (LLAGNs), 
and black hole X-ray binaries (BHXBs) in the hard and quiescent 
states (for more details see Narayan 2005; Yuan 2007; Ho 2008; 
Narayan \& McClintock 2008; Yuan 2011; Yuan \& Narayan 2014).

In recent years, many numerical hydrodynamic (HD) and 
magnetohydrodynamic (MHD) simulations have been performed
to study the structure and the dynamics of hot accretion flows 
(e.g., Igumenshchev \& Abramowicz 1999, 2000; Stone et al. 1999; 
Hawley et al. 2001; De Villiers et al. 2003; Igumenshchev et al. 2003; 
Yuan \& Bu 2010; Pang et al. 2011; Yuan et al. 2012a, 2012b; 
Narayan et al. 2012; Sadowski et al. 2013;
Bu et al. 2013; Yuan et al. 2015; Bu et al. 2016a,b). 
One of the most important and interesting findings, unlike pioneer 
analytical works that have been done on hot accretion flow, are
those simulations revealed that mass inflow rate is not constant 
and decreases inwardly which results the existence of
wind/outflow from system. It should be noted here that the properties 
and dynamics of the hot accretion flows with outflow is beyond 
the scope of this present work and we will postpone this hot 
topic to our future investigations.

Based on the self-similar assumption, many analytical works have 
also been done to investigate the structure and properties of hot 
accretion flow in one-dimension (e.g. Blandford \& Begelman 1999; 
Akizuki \& Fukue 2006; Abbassi et al. 2008; Zhang \& Dai 2008; 
Bu, Yuan \& Xie 2009; Mosallanezhad et al. 2012, Abbassi \& 
Mosallanezhad 2012a,b; Mosallanezhad et al. 2013) and also 
in two dimensions (e.g. Narayan \& Yi 1995a; Xu \& Chen 1997; 
Blandford \& Begelman 2004; Xue \& Wang 2005; Tanaka \& Menou 2006; 
Jiao \& Wu 2011; Mosallanezhad et al. 2014; Samadi \& Abbassi 2016; 
Mosallanezhad et al. 2016; Samadi et al. 2017). It is mentioned here, since the hot 
accretion flows are geometrically thick, the height-integrated approximation 
used in one dimensional self-similar solutions is not so appropriate. 
This is mainly because in this case, the physical variables are not only 
a function of $ r $ but also the function of vertical direction, $ \theta $, and
therefore solving the hot accretion flow in two dimensions is more reasonable. 
In all above mentioned works in two dimensions, their results only belong 
to the simplest case when the advection parameter, i.e., $ f $, was 
constant which may not be accurate. 

To answer to this question, how the advection parameter varies in vertical direction, 
several theoretical works have been done recently (see Gu et al. 2009; 
Samadi et al. 2014; Gu 2015; Zeraatgari \& abbassi 2015 (Hereafter ZA15)). 
For instance, ZA15 adopted polytropic relation in vertical direction instead of using 
energy equation. They also considered the modified “$ \alpha $” 
description of viscosity defined by Bisnovatyi-Kogan \& Lovelace 2007. 
By some modifications, they found an analytical solution for 
hot accretion flows and their results were totally in good agreement
with those presented in Narayan \& Yi 1995a without any difficulty 
of solving ordinary differential equations with two boundary conditions. 

In almost all above works magnetic field has not been included. 
It is now well known that magnetic field must be present and plays a 
significant role in the structure and the dynamics of the hot accretion flow, 
such as the angular momentum transfer by magnetorotational instability 
(MRI, Balbus \& Hawley 1998), the convective instability of the accretion flow 
(Narayan et al. 2012; Yuan et al. 2012b), and the driving mechanism of 
the wind/outflow (Yuan et al. 2015). Therefore, the aim of this paper is to 
consider the toroidal component of the magnetic field to our previous work, 
i.e., ZA15, and solve the flow equations including induction equation. 
The second change compare to our previous work is also adopting the 
modified $ \alpha $ description of viscosity for both the viscosity and 
the magnetic diffusivity due to the MRI.

The presented paper is structured as follows. In section 2, the basic equations and 
assumptions will be introduced. The self-similar solutions and boundary 
conditions are given in section 3. In section 4, numerical results will be 
presented. In section 5, we will summarize and conclude.

\section{Basic equations and Assumptions} \label{sec:equations}

In this section, we describe the basic equations for optically thin 
black hole accretion flows incorporating magnetic fields. We adopt 
spherical coordinates $ (r, \theta, \phi) $. The resistive MHD 
equations including conservation of mass, momentum, energy, 
and induction equation are as follows,

\begin{equation} \label{continiuty1}
  \frac{\partial \rho}{\partial t} + \nabla \cdot \left( \rho \mathbf{v} \right) = 0,
\end{equation}
\begin{equation} \label{momentum1}
    \rho  \left[ \frac{\partial \mathbf{v}} {\partial t} + \left(  \mathbf{v} \cdot \nabla  \right) \mathbf{v} \right] = 
    -\rho \nabla \psi - \nabla p + \nabla \cdot \mathbf{T} + \frac{\mathbf{J} \times \mathbf{B}}{c},
\end{equation}

\begin{equation} \label{energy1}
q_{\text{adv}} = q_{+} - q_{-} \equiv f q_{+},
\end{equation}

\begin{equation} \label{induction1}
    \frac{\partial \mathbf{B}}{\partial t} = \nabla \times \left( \mathbf{v} \times \mathbf{B} 
    - \frac{4 \pi}{c} \eta_{\text{m}} \mathbf{J} \right),
\end{equation}
where $ \rho $ is the density, $ \mathbf{v} $ is the velocity, $ p $ is 
the gas pressure, $ \psi $ is the gravitational potential of the central 
black hole, $ \mathbf{T} $ is the viscous stress tensor, $ \mathbf{B} $ 
is the magnetic field, $ \mathbf{J} = c \nabla \times \mathbf{B} /4 \pi $ 
is the current density, and $ \eta_{\text{m}} $ is the magnetic diffusivity. 
In the energy equation, $ q_{\text{adv}} $ is the advective cooling rate, 
$ q_{+} $ is the heating rate, $ q_{-} $ is the radiative cooling rate, and 
$ f $ represents the advection parameter which measures the fraction 
of the advection energy stored as entropy. We decomposed the heating 
rate into two components,

\begin{equation}\label{heating rate}
  q_{+} =  q_{\text{res}} + q_{\text{vis}}, 
\end{equation}

where, $ q_{\text{res}} $ and $ q_{\text{vis}} $  show heatings by dissipation of magnetic field
and viscosity, respectively. Based on numerical
simulation of Stone et al. 1999, we assume that the azimuthal component 
of the viscous tensor $ \mathbf{T} $ is the only non-zero component and 
is described as,

\begin{equation}
T_{r\phi} = \rho \nu r \frac{\partial}{\partial r} \left( \frac{v_{\phi}}{r} \right),
\end{equation}
where $ \nu $ is the kinematic viscosity. Heatings by dissipation of magnetic field
and viscosity can be written as,
\begin{equation}\label{q_res}
	q_{\text{res}} = \frac{4 \pi}{c^{2}} \eta_{\text{m}} \mathbf{J}^{2},
\end{equation}
\begin{equation}\label{q_vis}
	q_{\text{vis}} =  T_{r \phi} r \frac{\partial}{\partial r} \left(  \frac{v_{\phi}}{r} \right).
\end{equation}

In order to avoid the disparateness in terms of the turbulent viscosity and 
magnetic diffusivity, following Lovelace et al. 2009, we adopt the modified 
$ \alpha $ description of viscosity which in this model viscosity is not constant (see also Penna et al. 2013). We also assume both the viscosity and 
the magnetic diffusivity are due to the MRI as,
\begin{equation}\label{nu}
\nu = \mathcal{P} \eta_{\text{m}}  = \alpha \frac{p}{ \rho \Omega_{\text{K}}} g(\theta).
\end{equation}

Here, $ \mathcal{P} $ is the magnetic Prandtl number, $\alpha $ is the viscosity
parameter,  $ \Omega_{\text{K}} = (GM/r^{3})^{1/2} $ is the Keplerian angular 
velocity of the disk, and $ g(\theta) $ is a dimensionless function equal to unity 
and zero in the body and surface of the disk, respectively, (see Lovelace et al. 2009
; ZA15; Habibi et al. 2016). For simplicity, here we consider $  g(\theta) = \sin \theta $ 
in order to satisfy the above mentioned conditions. 
Following global and local 
MHD simulations of black hole accretion disks, the magnetic field inside the 
disk is decomposed into a large scale field and a fluctuating field. The effects 
of the fluctuating magnetic field  in transferring the angular momentum and 
dissipating the energy are described through the usual $ \alpha $ description 
(see $ \nabla \cdot \mathbf{T} $ in equation (\ref{momentum1}) and $ q_{\text{vis}} $ 
in equation (\ref{heating rate})).

As mentioned above, $ \mathbf{B} $ in equations (\ref{momentum1}) and 
(\ref{induction1}) corresponds to the large scale component of magnetic
field and we consider that the toroidal component is the only non-zero component of
magnetic field, $ \mathbf{B} = (0, 0,B_{\phi}) $. To solve the set of equations  
(\ref{continiuty1})-(\ref{induction1}), we assume a steady-state 
$ (\partial /\partial t = 0)$ and axisymmetric $ (\partial/ \partial \phi = 0) $ 
accretion flow. The gravitational potential of the central black hole is 
described in terms of the Newtonian potential, $ \psi = -(GM )/r $. Following 
Narayan \& Yi 1995a, we assume $ v_{\theta} = 0 $, which corresponds to a hydrostatic 
equilibrium in the vertical direction. However, this assumption may not be 
so appropriate when we want to investigate the effects of outflow on the dynamics 
of the accretion flow (see, e.g., Jiao \& Wu 2011; Mosallanezhad et al. 2014, 2016 
for more details). Therefore, the continuity equation, the three components 
of momentum equation, and also the induction equation are as follows,

\begin{equation} \label{continiuty2}
	\frac{1}{r^2} \frac{\partial}{\partial r} (r^2 \rho v_{r}) = 0,
\end{equation}

\begin{equation} \label{momentum_r2}
	\rho \left[ v_{r} \frac{\partial v_{r}}{\partial r} - \frac{v_{\phi}^2}{r} \right] = 
	-\rho \frac{GM}{r^2} - \frac{\partial p}{\partial r}
	+ \frac{1}{4\pi} \left(J_{\theta} B_{\phi}\right),
\end{equation}

\begin{equation} \label{momentum_t2}
	\rho \frac{v_{\phi}^{2}}{r} \cot\theta = \frac{1}{r}\frac{\partial p}{\partial \theta} 
	+ \frac{1}{4\pi}\left(J_{r} B_{\phi} \right),
\end{equation}

\begin{equation}\label{momentum_p2}
	\rho \left[ v_{r} \frac{\partial v_{\phi}}{\partial r} + \frac{v_{\phi} v_{r}}{r}  \right] = 
	\frac{1}{r^3} \frac{\partial}{\partial r} (r^3 T_{r\phi}),
\end{equation}

\begin{equation}\label{induction_2}
 	-\frac{\partial}{\partial r} (r v_{r} B_{\phi}) + \frac{\partial}{\partial \theta} (\eta J_{r})
 	- \frac{\partial}{\partial r} (r\eta J_{\theta}) = 0,
\end{equation}
where the components of current density, $ (\mathbf{J}) $, reads

\begin{equation}\label{Jr}
  J_{r} = \frac{1}{r \sin \theta} \frac{\partial}{\partial \theta} \left( B_{\phi} \sin \theta \right),
\end{equation}

\begin{equation}\label{Jt}
  J_{\theta} = - \frac{1}{r} \frac{\partial}{\partial r} \left( r B_{\phi} \right), 
\end{equation}

\begin{equation}\label{Jt}
  J_{\phi} = 0.  
\end{equation}

In this paper, we are interested in investigating the variation of the
advection parameter, $ f $,  with the polar angle in the case 
of hot accretion flows. Therefore, following Gu et al. 2009, 
Gu 2015, and ZA15, instead of using
energy equation, we apply the polytropic relation, $ p = K \rho^{\Gamma} $, 
in the $ \theta $ direction as our last equation. We also point out that 
some simulations revealed that the power index $ \Gamma $ is 
a constant less than unity (see e.g., Figure (3) of De Villiers et al. (2005)). 
This means that the time-averaged density drops faster than pressure 
from the equatorial plane to the rotation axis. Based on these 
results, we set $ \Gamma $ to be less than one throughout this paper.

\section{Self-similar solutions and boundary conditions} \label{sec:solutions}
\subsection{Self-Similar Solutions}
To better understanding the physics of hot accretion flow incorporating 
toroidal component of magnetic field, in this section, we seek self-similar 
solutions of the aforementioned equations. Self-similar solutions can 
describe the physical behavior of the accretion flow in an intermediate 
region of the disk (far away from inner and outer radial boundaries), and 
we think self-similar solutions are still good enough to study the variation of 
the physical variables within the disk. 

In plasma physics, parameter $ \beta $ measures the strength of the magnetic 
field in the plasma. In this part, we seek self-similar solutions along the vertical 
direction and represent the results in the case of $ \beta(\theta) $. 
The standard definition of plasma $ \beta $ is expressed as, 

\begin{equation} \label{PlasmaBeta}
	\beta \equiv \frac{p_{\text{gas}}}{p_{\text{mag}}}.
\end{equation}

Here, $ p_{\text{mag}} $ represents the magnetic pressure and because 
toroidal component is the only component of the magnetic field, the total
magnetic pressure is then given as,

\begin{equation} \label{PlasmaBeta}
 p_{\text{mag}} = \frac{B_{\phi}^{2}}{8\pi}.
\end{equation}

The radial self-similar solutions can be written as,

\begin{equation}
\rho (r,\theta) = \rho (\theta) r^{-3/2},
\end{equation}

\begin{equation}
v_{r}(r,\theta) =   r \Omega_{\text{K}}(r) v_{r}(\theta),
\end{equation}

\begin{equation}
v_{\phi}(r,\theta) = r \Omega_{\text{K}}(r)  \Omega(\theta) \sin \theta,
\end{equation}

\begin{equation}
p(r,\theta) = GM p(\theta) r^{-5/2},
\end{equation}

\begin{equation}
B_{\phi}^{2}(r,\theta) = 4 \pi GM b(\theta)^{2} r^{-5/2}.
\end{equation} 
Substituting the above self-similar assumptions into the equations 
of system, and using polytropic relation in the vertical direction, we 
can rewrite the main equations as,
\begin{multline} \label{momentum_r3}
\rho(\theta) \left[ -\frac{1}{2} v_{r}(\theta)^2 - \Omega(\theta)^{2} \sin^{2} \theta \right] = 
- \rho(\theta) + \frac{5}{2} p(\theta)  \\ 
+ \frac{1}{4} b(\theta)^{2},
\end{multline}

\begin{multline}\label{momentum_t3}
 \rho(\theta) \Omega(\theta)^{2} \sin \theta \cos \theta =  \frac{dp(\theta)}{d\theta} 
+ b (\theta) \frac{db(\theta)}{d\theta} \\
+ b(\theta)^{2} \cot \theta, 
\end{multline}

\begin{equation}\label{momentum_p3}
	 \rho(\theta) v_{r}(\theta) = - \frac{3}{2} \alpha p(\theta) \sin \theta,
\end{equation}

\begin{equation} \label{dimensionless_polytropic}
    \frac{dp(\theta)}{d \theta} = K \Gamma  \rho(\theta)^{\Gamma - 1} \frac{d \rho(\theta)}{d \theta},
\end{equation}

\begin{multline} \label{induction_variable}
	\frac{d^{2}b(\theta)}{d \theta^{2}} =  \left[ \frac{9}{8} \mathcal{P} - \frac{3}{16} \right] b(\theta)  + b(\theta) \csc^{2}\theta - \frac{db(\theta)}{d \theta} \cot \theta  \\
	+ \left( \frac{d b(\theta)}{d \theta}  +  b(\theta) \cot \theta \right)  \left( \frac{d \ln \rho(\theta)}{d \theta} - \frac{d \ln p(\theta)}{d \theta} - \cot \theta \right). 
\end{multline}
Now, we have a set of ordinary differential equations for given values of $ \alpha, K, \Gamma, \beta $, and $ \mathcal{P} $
that should be solved numerically with the boundary conditions will be introduced in the next subsection.

\subsection{Boundary Conditions}

Equations (\ref{momentum_r3})-(\ref{induction_variable}) are differential 
equations for three variables: $ \rho(\theta) $, $ b(\theta) $, and $ db(\theta)/d\theta $. 
Actually, other variables such as $ p(\theta) $, $ v_{r}(\theta) $, and 
$ v_{\phi}(\theta) [= \Omega(\theta) \sin \theta] $ can be determined by our 
main three variables. In this work, we assume that the accretion flow is 
evenly symmetric about the mid-plane, i.e., $  \rho(\theta) = \rho(\pi - \theta) $, 
$ p(\theta) = p (\pi - \theta) $, $ v_{r}(\theta) = v_{r} (\pi - \theta) $,
$ v_{\phi} (\theta) = v_{\phi} (\pi - \theta) $, and $ b(\theta) = - b(\pi - \theta) $.
At the mid-plane of the disk we have by the symmetry,

\begin{equation} \label{symmetry}
	\theta = \frac{\pi}{2} \qquad \qquad  \frac{d p}{d \theta} = 
		     \frac{d \rho}{d \theta} = \frac{d b}{d \theta} = 0.
\end{equation}

In essence, to solve equations (\ref{momentum_r3})-(\ref{induction_variable}),
besides the symmetric boundary conditions at $ \theta = 90^{\circ} $,
it is required to apply appropriate boundary conditions at the rotation axis, i.e.,
$ \theta = 0^{\circ} $. Indeed, this is a two-point boundary value problem
which the solutions behave properly in the whole $ r-\theta $ space.
We have tried to obtain such a solution but failed. This is a caveat in this work.  
Here, in contrast to our previous work, i.e., Zeraatgari \& Abbassi (2015), 
we shoot from $ \theta = \pi/2 $ towards the axis and simply stop 
the integration when we meet unphysical solution.
Therefore, we only require the solution satisfying the boundary condition at $ \theta = \pi/2 $.
We think the solution obtained in this way should still be physically meaningful.

As a boundary condition, we set the density to be $ \rho(\pi/2) = 1.0 $ 
at the equatorial plane throughout this paper. Under the above boundary conditions and 
symmetries, equation (\ref{induction_variable}) can be simplified 
into the following equation,

\begin{equation} \label{db_90}
	\frac{d^{2}b(\pi/2)}{d \theta^{2}}  = \frac{1}{8} \left[ 9 \mathcal{P} + \frac{13}{2} \right] b(\pi/2).
\end{equation}

It should be noted here that since the second derivation of 
toroidal component of magnetic field has positive value at
the mid-plane of the disk therefore, this clearly implies that 
$ b(\pi/2) $ should be minimum there. We adopt the 
values of $ b(\pi/2) $ with reference to our previous study 
(see Mosallanezhad et al. 2014, 2016). We do find that by 
considering any small value for the toroidal component of 
the magnetic field at the beginning of the integration,
at a certain critical value of $ \theta $, denoted as 
$ \theta_{s} $, $ v_{\phi}^{2} (\theta_s)$ begins to vanish and 
becomes zero. Then, at the region of $ \theta < \theta_{s} $, 
$ v_{\phi}^2 (\theta) $ becomes negative and the solution is no longer physical. 

The solution in the region of $ \theta_s < \theta < \pi/2 $ is still physical, mainly because, on the one hand, the solution satisfies the equations and the boundary conditions at $ \pi/2 $ and on the other hand, the values of all physical quantities of the accretion flow at $ \theta_s $ are also physical. Therefore, we can reasonably treat these values as boundary conditions at $ \theta_s $. Another caveat that needs to be mentioned
here is that we consider constant values for $ K $, $ \rho $ 
and $ b $ at the equator. As a result, the pressure and sound speed are constants there and would 
not be self-consistently determined as part of the solution.

\section{numerical results}  \label{sec:Results}

\begin{figure*}
\includegraphics[width=\textwidth,angle=0]{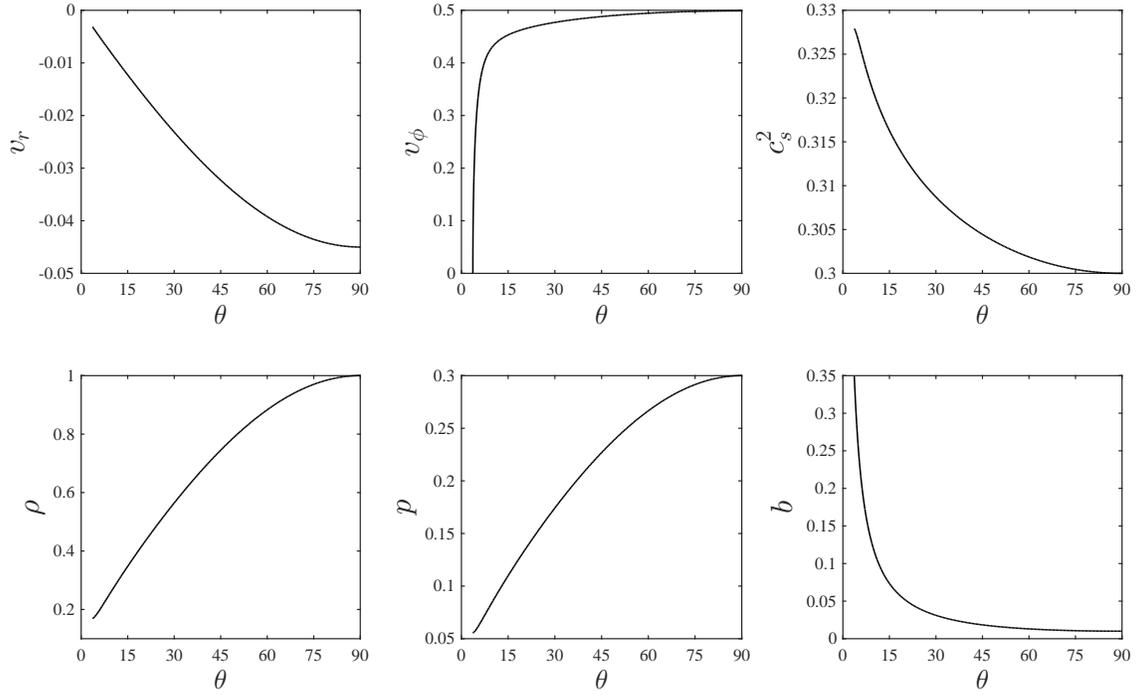}
\caption{Angular profiles of physical variables. 
Here $ \alpha = 0.1 , \mathcal{P} = 1.0, K = 0.3,  \Gamma = 0.95 , b(90) = 10^{-2},\text{and}\  \rho(90) = 1.0 $.}
\label{fig:variables}
\end{figure*}
\begin{figure*}
\includegraphics[width=\textwidth,angle=0]{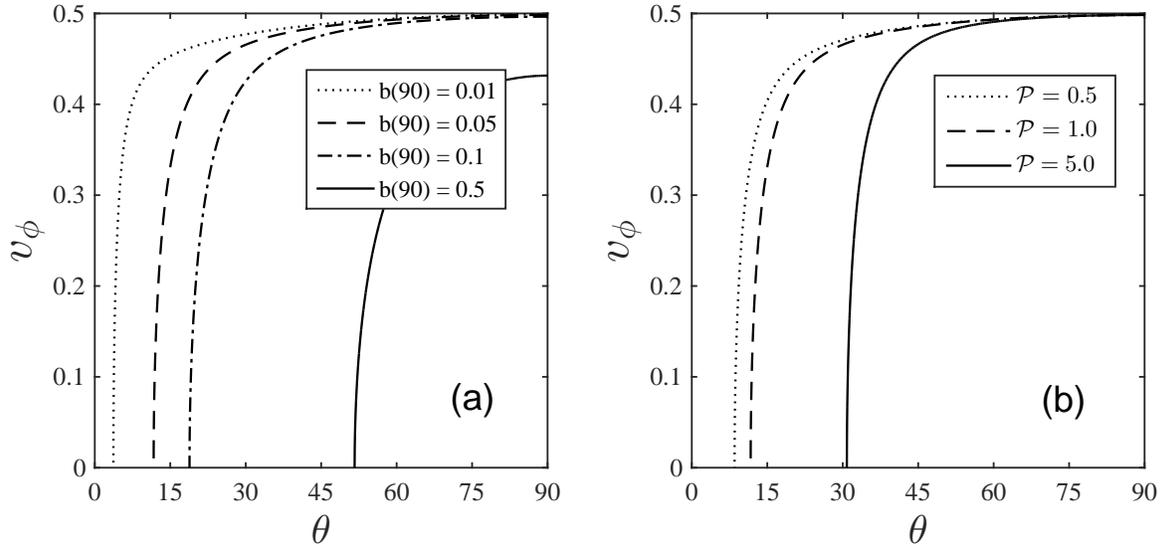}
\caption{Angular profiles of azimuthal velocity for (a): different values of magnetic field strength and (b): different Prandtl numbers.
Here $ \alpha= 0.1 , \mathcal{P} = 1.0, K = 0.3,  \Gamma = 0.95,\text{and}\ b(90) = 10^{-2}.$ }
\label{fig:vp}
\end{figure*}
%


\begin{figure}
\includegraphics[width=95mm,angle=0]{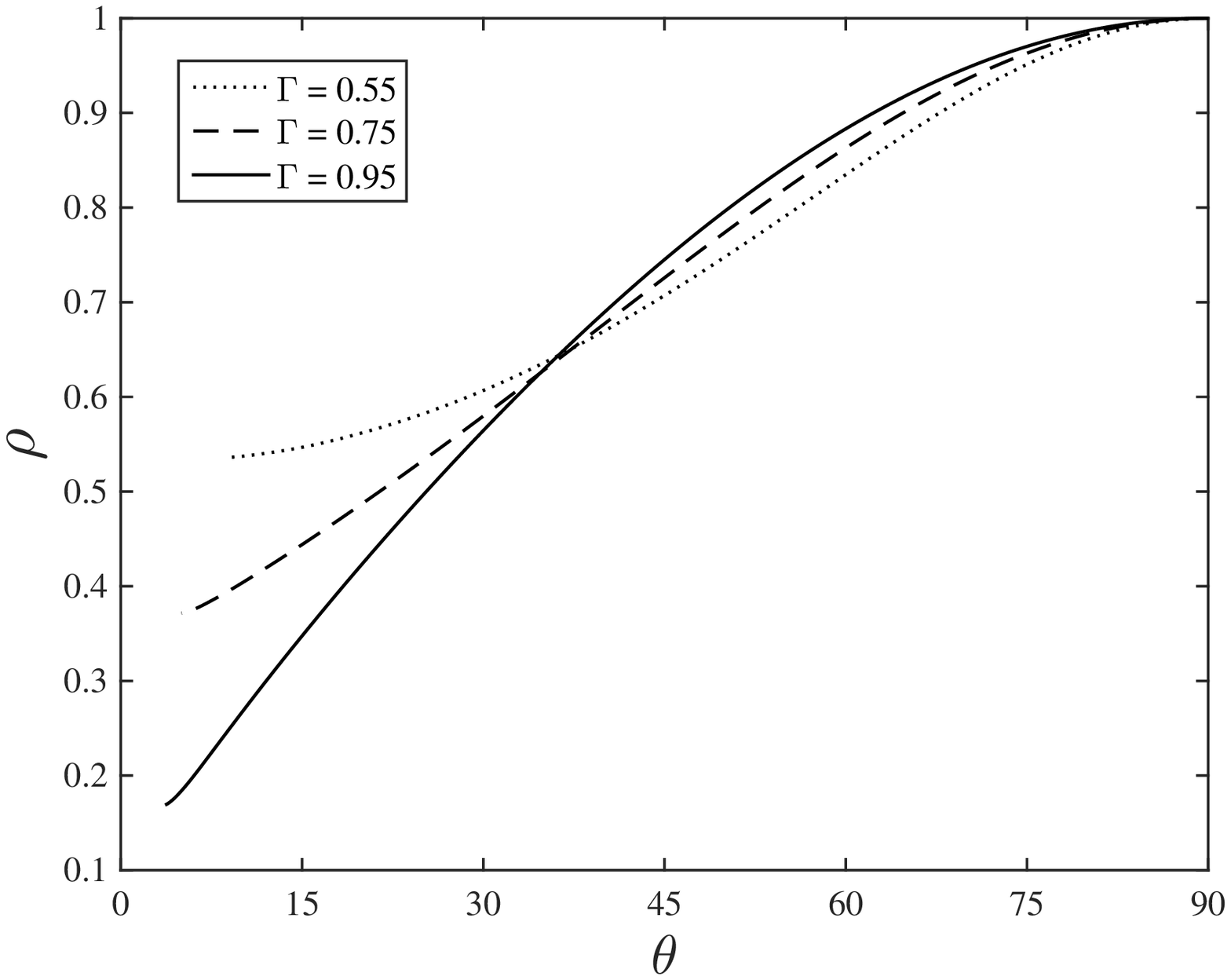}
\caption{Angular profile of density for given values of $ \Gamma $.
Here $ \alpha = 0.1,  \mathcal{P} = 1.0, K = 0.3, b(90) = 10^{-2},\text{and}\ \rho(90) = 1.0 $.}
\label{fig:rho}
\end{figure}

\begin{figure*}
\includegraphics[width=\textwidth,angle=0]{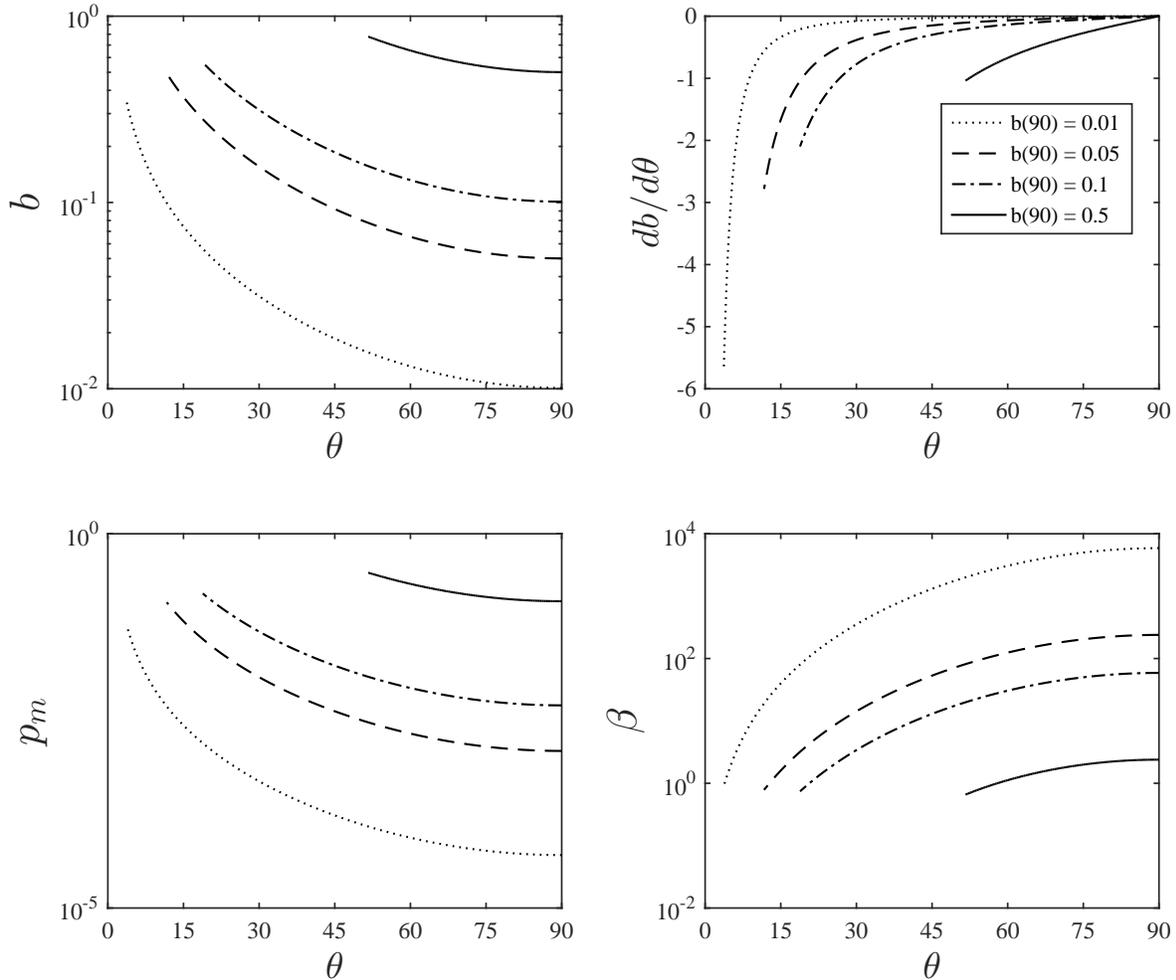}
\caption{Angular profiles of magnetic field variables for given values of magnetic field strength at the equatorial plane. Here $ \alpha = 0.1,  \mathcal{P} = 1.0, K = 0.3, \Gamma = 0.95,\text{and}\ \rho(90) = 1.0 $.}
\label{fig:b90}
\end{figure*}
\begin{figure*}
\includegraphics[width=\textwidth,angle=0]{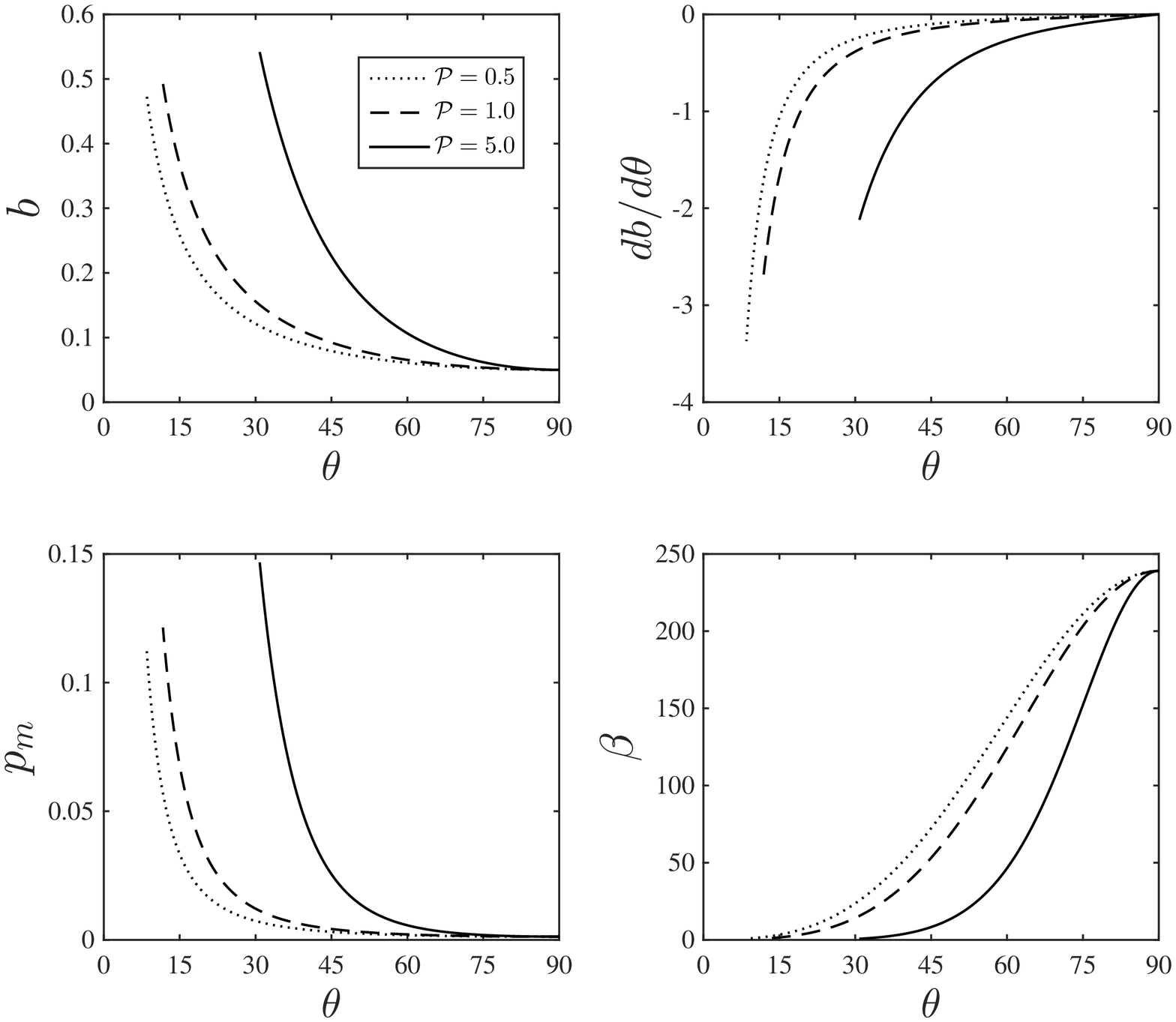}
\caption{Angular profiles of magnetic field variables for given values of prandtl number. Here $ \alpha = 0.1,  \mathcal{P} = 1.0, K = 0.3, \Gamma = 0.95, b(90) = 0.05,\text{and}\ \rho(90) = 1.0 $.}
\label{fig:pr}
\end{figure*}
\begin{figure*}
\includegraphics[width=\textwidth,angle=0]{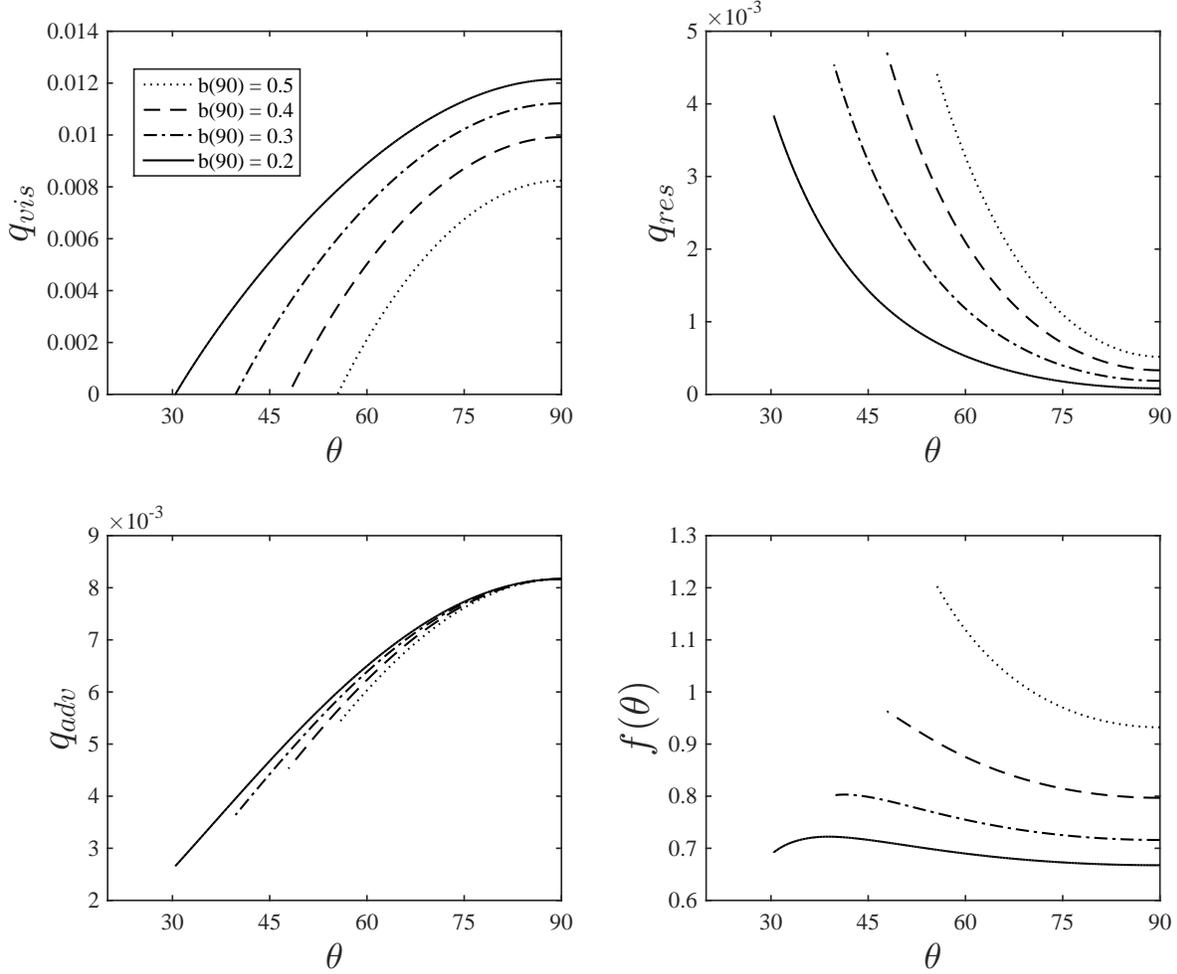}
\caption{Cooling and heating for different values of magnetic field strength at the equatorial plane, top-left: viscous heating, top-right: resistive heating, bottom-left: advective cooling, and bottom-right: advection parameter. Here $ \alpha = 0.1, \mathcal{P} = 1.0
, \Gamma = 0.95,\text{and}\ K = 0.33 $.}
\label{fig:qs}
\end{figure*}
\begin{figure}
\includegraphics[width=90mm,angle=0]{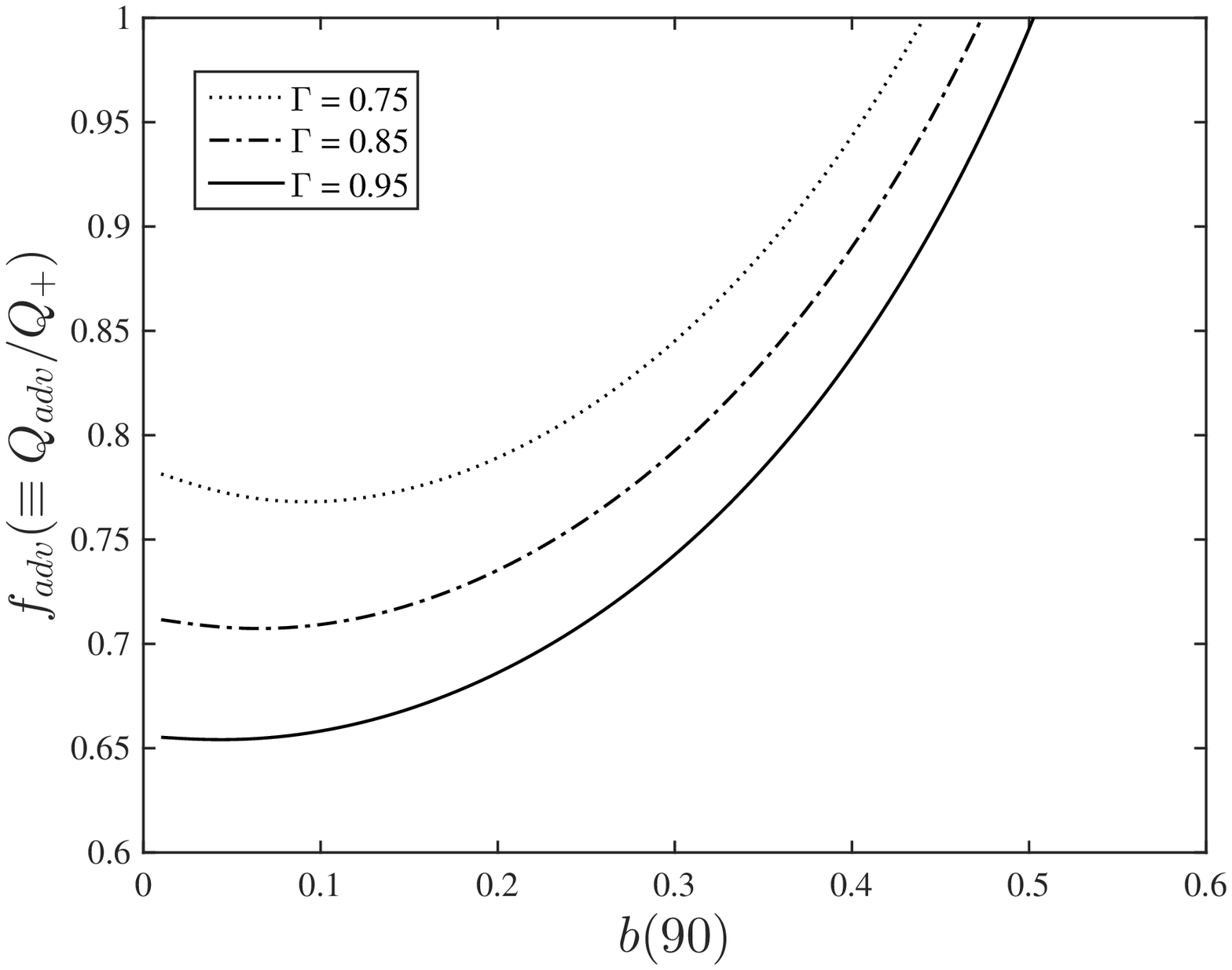}
\caption{The variation of energy advection parameter versus the strength of the magnetic field at the equatorial plane for given values of $ \Gamma $.
Here $ \alpha = 0.1,  \mathcal{P} = 1.0, K = 0.33 $.}
\label{fig:fadv}
\end{figure}

We have solved the equations (\ref{momentum_r3})-(\ref{induction_variable}) 
numerically. In all of solutions $ \alpha = 0.1 $. The results have been shown 
in Figures (\ref{fig:variables})-(\ref{fig:fadv}). Figure (\ref{fig:variables}) shows 
the effect of magnetic field on angular profiles of physical variables from 
equatorial plane ($ \theta = 90^{\circ} $) to rotation axis ($ \theta = 0^{\circ} $), 
for K = 0.3,  $ \Gamma = 0.95 $ , $ b(90) = 10^{-2} $, and  $ \rho(90) = 1.0 $. 
From left to right and top to bottom the plots of $ v_{r} $, radial velocity; 
$ v_{\phi} $, azimuthal velocity; $ c_{s}^2 $, sound speed squared; $ \rho $, 
density; $ p $, pressure; and $ b $, toroidal component of magnetic field have been presented, respectively. 
It is clear, the radial and azimuthal components of velocity decrease 
towards the rotation axis. As we explained in previous section, at a certain 
angle, the azimuthal component of velocity is null. 
Therefore, we stop the integration there and consider this angle as the surface of the disk. 
Figure (\ref{fig:variables}) also shows sound speed squared which represents 
the disk temperature increases and reaches to a value near viral temperature 
near the disk surface. Moreover, since the density has been normalized to unity 
at the equatorial plane decreases towards the rotation axis. This clearly shows that 
the maximum accretion process happens in a regime near the disk mid-plane 
which is in good agreement with those obtained from numerical simulations. 
Also, at the equatorial plane the pressure is the maximum and decreases 
towards the rotation axis. Compare to our previous paper (see, Figure (2) of ZA15), 
magnetic field is a new variable here, and the toroidal component of the magnetic field 
has the minimum value at the mid-plane and increases towards the surface of the disk. 
This is mainly because the second derivative of the toroidal component of the magnetic 
field has positive sign at the equatorial plane of the disk (see Eq. (\ref{db_90})), 
which means there should exist a minimum for the magnetic field there. These results all are in 
agreement with previous analytical works done in the presence of magnetic field (see e.g., 
Mosallanezhad et al. 2014, 2015 and Samadi \& Abbassi 2014). 

As it expressed in the introduction, the main purpose of this work is to study 
the effects of the toroidal component of the magnetic field on the structure and 
the dynamics of the hot accretion flows. To do so, first we investigate how magnetic 
field changes the disk surface. Here, two parameters show the effects of 
magnetic field including $ b(90) $ which is the value of the toroidal component 
of the magnetic field at the equatorial plane of the disk and also the magnetic 
Prandtl number, $ \mathcal{P} $. Figure (\ref{fig:vp}) represents the angular 
profile of the azimuthal velocity for different values of the magnetic field strength at the equatorial 
plane (panel a), and different Prandtl numbers (panel b). In panel a, the dotted, 
dashed, dot-dashed, and solid lines are for $ b(90) = 0.01, 0.05, 0.1 $, and $ 0.5 $, 
respectively. It is clear that azimuthal velocity decreases with increasing 
the magnetic field strength at the equatorial plane. For minimum value of $ b(90) $, i.e., 
$ b(90) = 0.01 $, the disk surface is around $ \sim 5^{\circ} $ which is very close to 
the rotation axis while for the higher value of the magnetic field strength, $ b(90) = 0.5 $, azimuthal velocity becomes 
null near $\sim  53^{\circ} $. In panel b, the dotted, dashed, and solid lines are for 
$ \mathcal{P} = 0.5, 1.0 $, and $ 5.0 $, respectively. It should be noted here that 
based on equation (\ref{nu}),  since we fix $ \alpha = 0.1 $ throughout this paper 
therefore, those values for Prandtl number correspond to $ \eta_{\text{m}} = 0.05, 
0.1, 0.5 $, respectively. It can be seen clearly that the increasing the Prandtl 
number does reduce the azimuthal velocity. For $ \mathcal{P} = 0.5 $, the disk 
surface is around $\sim 10^{\circ}  $ and for $ \mathcal{P} = 5.0 $, the disk surface
is approximately $ \sim 30^{\circ} $. Therefore, we conclude that in the 
case of no-outflow solution, the toroidal component of the magnetic field can decrease
the disk surface and in comparison with non-magnetized hot accretion disks, magnetized 
disks might be thiner. 


Figure (\ref{fig:rho}) is angular profile of density for three values of $ \Gamma $. 
As we explained in previous sections, we set the value of $ \Gamma $ to be less 
than unity. This figure shows the variation of mass density for different 
values of $ \Gamma $, so we can check how this index can affect the density profile. 
The dotted, dashed, and solid lines are for $ \Gamma = 0.55, 0.75 $, and $ 0.95 $, respectively.  
The mass density from equatorial plane towards the rotation axis increases 
with $ \Gamma $ around $ \theta = 40^{\circ} $. As you can see, for $ \theta \lesssim 40^{\circ} $, around rotation 
axis, the variations of the density with respect to $ \Gamma $ becomes inverse so, it has a decreasing trend. 
This result clearly shows that the density rapidly decreases for higher values of 
$ \Gamma $ from equatorial plane towards the disk surface. Therefore, from now 
we set $ \Gamma $ to be nearly unity because of two main reasons. 
First, numerical simulations of hot accretion flows show that the mass density rapidly 
decreases near the disk surface (see, e.g., Villiers et al. 2005; Yuan \& Bu 2010), 
and second, we want to compare this results with our previous HD case (ZA15).    

Figure (\ref{fig:b90}) shows angular profiles of magnetic field variables 
including $ b $, toroidal component of magnetic field, $ db/d\theta $, the first derivation of magnetic 
field, $ p_{\text{m}} $, magnetic pressure, and $ \beta(\theta) $, plasma beta.
All variables are plotted for four values of magnetic field strength
at the equatorial plane. Dotted, dashed, dot-dashed, and solid lines are for 
$ b(90) = 0.01, 0.05, 0.1 $, and $ 0.5 $, respectively. Clearly, with increasing 
the strength of magnetic field at the equatorial plane, $ b $ increases in allover the flow. 
Also, as it is shown in top-right panel, $ db/d\theta $ becomes increasingly negative towards the disk surface 
and is null at the disk mid-plane 
. In bottom-left panel magnetic pressure, 
$ p_{m} $, has an increasing trend from $ \theta=90^{\circ} $ towards the disk surface. 
This is mainly because the dimensionless magnetic pressure is proportion to
$ b^{2} $ and an increase in the magnetic field strength at the equatorial plane will 
cause the magnetic pressure increases. In bottom-right panel, it is clear that 
plasma beta, $ \beta(\theta) = p_{\text{gas}}/p_{\text{mag}}$, the ratio of 
gas pressure to magnetic pressure, decreases from equatorial plane towards 
the rotation axis and reaches to the value near unity at the disk surface. 
This is mainly because the gas pressure, the numerator, decreases from 
equatorial plane towards the rotation axis while magnetic pressure, the denominator, 
increases. In addition, when the magnetic field strength at the equatorial 
plane increases, $ \beta $ decreases. 

Figure (\ref{fig:pr}) shows angular 
profiles of magnetic field variables for given values of Prandtl number including 
$ \mathcal{P} = 0.5, 0.1 $, and $ 5 $. As you can see, as Prandtl number increases, 
$ b, db/d\theta $, and $ p_{m} $ increase, while $ \beta $ decreases, the same trend as we 
discussed in Figure (\ref{fig:b90}). These results are also in good agreement with 
those obtained from numerical simulations and analytical works 
(Mosallanezhad et al. 2014, 2015; Yuan \& Bu 2010; Yuan et al. 2012b).

As we described in previous section, we applied polytropic equation of state 
in vertical direction instead of energy equation to answer to this question, how 
the advection parameter, $ f $, varies in vertical direction? Moreover, since our numerical 
results are in good agreement with previous works done with fixed advection 
parameter ($ f = 1.0 $) such as Narayan \& Yi 1995a, Mosallanezhad et al. 2014, 
2016, Samadi \& abbassi 2016, we conclude that our assumptions 
and results are reliable. Now, we investigate the variation of advection 
parameter with magnetic field parameters. To do so, first we plot the 
heating/cooling terms along vertical direction before integrating the heating 
rates and advective cooling rate in energy equation with $ \theta $ angle. 
The top-left and top-right panels of Figure (\ref{fig:qs}) show angular profiles 
of viscous dissipation heating rate $ q_{\text{vis}} $ and magnetic field dissipation heating 
rate $ q_{\text{res}} $, respectively. In the top-left panel, viscous dissipation 
decreases from equatorial plane of the disk towards the rotation axis. 
Also, as $ b(90) $ increases, viscous dissipation decreases. In contrast, 
the angular profile of magnetic dissipation heating rate increases from 
equatorial plane towards the disk surface. Moreover, magnetic field dissipation 
heating increases with increasing magnetic field strength at the equatorial plane. 
The reason of these behavior can be easily understood from equations 
(\ref{q_res}) and (\ref{q_vis}), where $ q_{\text{vis}} \propto v_{{\phi}} $ and 
$ q_{\text{res}} \propto b^{2} $ (see Figure (\ref{fig:variables})). 
The bottom-left panel of Figure (\ref{fig:qs}) is angular profile of 
advective cooling, $ q_{\text{adv}} $, which decreases 
from the equatorial plane towards the rotation axis. 
Also, advective cooling decreases with an increase in the value of magnetic field strength 
at the equatorial plane of the disk.
The angular profile of advection parameter, 
$ f(\theta) (= q_{\text{adv}}/q_{+}) $, that is represented in bottom-right panel 
has an increasing trend from mid-plane of the disk towards the disk surface. 
In addition, with increasing the value of magnetic field strength at the equatorial plane of the disk,
the advection parameter will be increased.
As it can be seen, the solution with largest values of magnetic field 
at the equatorial plane, i.e., $ b(\pi/2) > 0.4 $ is likely unphysical. 
This is mainly because at some $ \theta $ angles the advection parameter 
will become greater than unity and it would rule out the solutions with this property.

The vertical integration of the viscous dissipation heating, magnetic field dissipation 
heating, and also advective cooling are as follows,

\begin{equation} \label{Qvis}
	Q_{\text{vis}} = 2 \int^{\pi/2}_{\theta_{s}} q_{\text{vis}}\ r \sin \theta \ d\theta,
\end{equation}

\begin{equation} \label{Qvis}
	Q_{\text{res}} = 2 \int^{\pi/2}_{\theta_{s}} q_{\text{res}}\ r \sin \theta \ d\theta,
\end{equation}

\begin{equation} \label{Qvis}
	Q_{\text{adv}} = 2 \int^{\pi/2}_{\theta_{s}} q_{\text{adv}}\ r \sin \theta \ d\theta.
\end{equation}

Then, the energy advection parameter can be defined as,
\begin{equation} \label{fadv}
	f_{\text{adv}} = \frac{Q_{\text{adv}} }{Q_{+}},
\end{equation}
where $ Q_{+} = Q_{\text{vis}}  + Q_{\text{res}}  $. Figure (\ref{fig:fadv}) shows 
the variation of energy advection parameter, $ f_{\text{adv}} $, versus 
the magnetic field strength at the equatorial plane for different values of 
$ \Gamma $. The dotted, dot-dashed, and solid lines are for $ \Gamma = 0.75, 0.85 $, 
and $ 0.95 $, respectively.  Here, $ \alpha = 0.1 $, $ \mathcal{P} = 1.0 $, $ K = 0.33 $, 
and $ \rho(90) = 1.0 $. It is seen that the energy advection parameter decreases as 
$ \Gamma $ increases. Moreover, energy advection parameter becomes unity for 
$ 0.4 < b(90) \leq 0.5  $. This result illustrates that advective cooling can 
balance total dissipation heating rate. Consequently, Our results also show that 
in terms of the no-outflow solution incorporating toroidal magnetic field, thermal 
equilibrium still exists for both strong magnetic filed at the equatorial plane 
($ b(90) \sim 0.5 $) and higher values of $ \Gamma $ index. 

\section{Summary and Conclusions}  \label{sec:Summery}
To summarize we have solved MHD equations of optically thin geometrically 
thick black hole accretion flows in spherical coordinates $ (r, \theta, \phi) $. 
The gravitational potential of the central black hole was assumed Newtonian. 
The cooling in the energy equation was considered advective cooling, and 
the heating rate was decomposed into two components, magnetic field and 
viscosity dissipations. We assumed both the viscosity and the magnetic diffusivity 
are due to MRI and adopted the modified α description of viscosity. The azimuthal 
component of the viscous stress tensor, $ T_{r\phi} $, was considered as the 
only non-zero component. We only considered the toroidal component of magnetic 
field, $ B_{\phi} $. We did not consider wind/outflow in the flow so, $ v_{\theta} = 0 $. 
Instead of using energy equation, we applied the polytropic relation in the vertical direction. 
Following some simulations revealed the power index $ \Gamma $ is a constant 
less than unity, we set $ \Gamma $ to be less than one throughout this paper. 
We used self-similar solutions to solve the MHD equations because it is good 
for studying the main body of the flow far from the inner and outer radial boundaries. 

We could not solve a two-point boundary value problem
therefore, we shifted to singular boundary value problem and would think that our solution is still 
physically meaningful.
We found that there is a surface for the flow at some angle, $ \theta_{s} $, 
where the azimuthal velocity, $ v_{\phi} $, is zero. This result is satisfied by any small 
value of the toroidal component of magnetic field at the beginning of the integration.

Our numerical results show that the radial and azimuthal components of velocity decrease 
towards the rotation axis while the sound speed squared increases. 
Moreover, other physical variables such as mass density and 
gas pressure drop rapidly towards the disk surface. Although, the
magnetic pressure has a minimum at the equatorial plane  and 
reaches to its maximum value around the disk surface. 
Consequently, the plasma beta, $ b(\theta) = p_{\text{gas}}/p_{\text{mag}} $, 
decreases towards the rotation axis and approximately reaches to unity at the disk surface. 
In spite of the simplicity of our model in viscosity, magnetic field and the disk itself, 
we think that the presented analytical results give us a better understanding of such a 
complicated system incorporating magnetic field. It is good to note here that our numerical 
results are in good agreement with some previous simulations and analytical studies 
(see, Yuan et al. 2012a; Yuan et al. 2015; Mosallanezhad 2014, 2016; Samadi \& Abbassi 2016).

We found that the viscous dissipation heating decreases from 
equatorial plane of the disk towards the rotation axis, and 
decreases for higher amounts of magnetic field strength at the equatorial 
plane of the disk. In contrast, the angular profile of magnetic dissipation heating 
increases from equatorial plane towards the disk surface.

Our main conclusion was that in terms of the no-outflow solution incorporating 
toroidal magnetic field, thermal equilibrium still exists for both 
strong magnetic filed at the equatorial plane of the disk
and for higher values of $ \Gamma $ index. 

As we mentioned in \ref{sec:intro}, magnetic field can transfer 
angular momentum by MRI which is related with the wind production. 
So, in our future work we will study the effect of the magnetic field 
on the dynamics and the structure of hot accretion flow extracting 
the wind from the system. For better understanding of the system, 
we will also include all components of velocity and magnetic field 
in MHD equations to check how global magnetic field affects the 
dynamics of hot accretion flows such as radial and angular velocity. 

\acknowledgements
This work is supported by National Natural Science Foundation of 
China (Grant No. U1431228, 11133005, 11233003, and 11421303).


\begin{thebibliography}{}


\bibitem[\protect\citeauthoryear{}{}]{}  Bu D., Yuan F., Wu M., Cuadra J., 2013, MNRAS, 434, 1692
\bibitem[\protect\citeauthoryear{}{}]{} Abbassi S., Ghanbari J., Najjar S., 2008, MNRAS, 388, 663
\bibitem[\protect\citeauthoryear{}{}]{} Abbassi, S., \& Mosallanezhad, A. 2012a, Ap\&SS, 341, 375\bibitem[\protect\citeauthoryear{}{}]{} Abbassi, S., \& Mosallanezhad, A. 2012b, RAA, 12, 1625
\bibitem[\protect\citeauthoryear{}{}]{} Abramowicz M. A., Czerny, B., Lasota, J. P., Szuszkiewicz, E., 1988, \apj,  58, 332
\bibitem[\protect\citeauthoryear{}{}]{} Abramowicz, M. A., Chen, X., Kato, S., Lasota, J. P., Regev, O., 1995, ApJL, 438, L37
\bibitem[\protect\citeauthoryear{}{}]{} Abramowicz, M. A., Czerny, B., Lasota, J. P., \& Szuszkiewicz, E., 1988, ApJ, 332, 646
\bibitem[\protect\citeauthoryear{}{}]{} Abramowicz, M. A., Fragile, P. C. 2013, Living Rev. Relativ., 16, 1
\bibitem[\protect\citeauthoryear{}{}]{} Akizuki C., Fukue J., 2006, PASJ, 58, 469
\bibitem[\protect\citeauthoryear{}{}]{} Balbus S. A., Hawley J. F., 1998, Rev. Mod. Phys., 70, 1
\bibitem[\protect\citeauthoryear{}{}]{} Begelman, M. C., 1979, MNRAS, 187, 237
\bibitem[\protect\citeauthoryear{}{}]{} Begelman, M. C., Meier, D. L., 1982, \apj, 253, 873
\bibitem[\protect\citeauthoryear{}{}]{} Bisnovatyi-Kogan, G. S., \& Lovelace, R. V. E. 2007, ApJ, 667, 167
\bibitem[\protect\citeauthoryear{}{}]{} Blaes, O. 2014, Space Science Reviews, 183, 21
\bibitem[\protect\citeauthoryear{}{}]{} Blandford R. D., Begelman M. C., 2004, MNRAS, 349, 68
\bibitem[\protect\citeauthoryear{}{}]{} Blandford, R. D., \& Begelman, M. C. 1999, MNRAS, 303, L1
\bibitem[\protect\citeauthoryear{}{}]{} Bu D.-F., Yuan F. ,Gan Z.-M., Yang, X.-H., 2016a, APJ, 818, Issue 1, article id 83,8 pp
\bibitem[\protect\citeauthoryear{}{}]{} Bu D.-F., Yuan F. ,Gan Z.-M., Yang, X.-H., 2016b, APJ, 823, Issue 2, article id 90,6 pp
\bibitem[\protect\citeauthoryear{}{}]{} Bu D., Yuan F., Xie F., 2009, MNRAS, 392, 325
\bibitem[\protect\citeauthoryear{}{}]{} Chen, X., \& Taam, R. 1993, \apj, 412, 254
\bibitem[\protect\citeauthoryear{}{}]{} De Villiers, J. P., Hawley, J. F., \& Krolik, J. H. 2003, ApJ, 599, 1238
\bibitem[\protect\citeauthoryear{}{}]{} De Villiers, J.-P., Hawley, J. F., Krolik, J. H., \& Hirose, S. 2005, ApJ, 620, 878
\bibitem[\protect\citeauthoryear{}{}]{} Eggum, G. E., Coroniti, F. V., \& Katz, J. I. 1988, \apj, 330, 142
\bibitem[\protect\citeauthoryear{}{}]{} Frank J, King A., Raine D. 2002, Accretion Power in Astrophysics, 3rd ed. (Cambridge University Press, Cambridge)
\bibitem[\protect\citeauthoryear{}{}]{} Fukue, J. 2004, PASJ, 56, 569
\bibitem[\protect\citeauthoryear{}{}]{} Gu, W.-M. 2015, ApJ, 799, 71\bibitem[\protect\citeauthoryear{}{}]{} Gu, W.-M., Xue, L., Liu, T., \& Lu, J.-F. 2009, PASJ, 61, 1313
\bibitem[\protect\citeauthoryear{}{}]{} Habibi, A., Abbassi, S., \& Shadmehri, M. 2016, MNRAS, 464, 5028
\bibitem[\protect\citeauthoryear{}{}]{} Hawley, J., Balbus, S. A., \& Stone, J. M. 2001, ApJ, 554, L49
\bibitem[\protect\citeauthoryear{}{}]{} Ho, L. 2008, ARA\&A, 46, 475
\bibitem[\protect\citeauthoryear{}{}]{} Ichimaru S, 1977, \apj, 214, 840
\bibitem[\protect\citeauthoryear{}{}]{} Igumenshchev, I. V., \& Abramowicz, M. A., 1999, MNRAS, 303, 309
\bibitem[\protect\citeauthoryear{}{}]{} Igumenshchev, I. V., \& Abramowicz, M. A., 2000, ApJS, 130, 463
\bibitem[\protect\citeauthoryear{}{}]{} Igumenshchev, I. V., Narayan, R., \& Abramowicz, M. A. 2003, ApJ, 592, 1042
\bibitem[\protect\citeauthoryear{}{}]{} Jiao C. L., Wu X. B., 2011, ApJ, 733, 112
\bibitem[\protect\citeauthoryear{}{}]{} Kato, S., Fukue, J.,\& Mineshige, S. 2008, Black-HoleAccretionDisks: Towards a New Paradigm (Kyoto: Kyoto Univ. Press)
\bibitem[\protect\citeauthoryear{}{}]{} Kats, J. I. 1977, ApJ, 215, 265
\bibitem[\protect\citeauthoryear{}{}]{} Lasota, J. P. 2016, Astrophysics and Space Science Library, In press. (arXiv:1505.02172)
\bibitem[\protect\citeauthoryear{}{}]{} Lovelace, R. V. E., Bisnovatyi-Kogan, G. S., \& Rothstein, D. M. 2009, NPGeo, 16, 77
\bibitem[\protect\citeauthoryear{}{}]{} Lynden-Bell, D., Pringle, J. E., 1974, MNRAS, 37, 168
\bibitem[\protect\citeauthoryear{}{}]{} Mosallanezhad, A., Abbassi, S., Shadmehri, M., Ghanbari, J., 2012, Ap \& SS, 337, 703M
\bibitem[\protect\citeauthoryear{}{}]{} Mosallanezhad, A., Abbassi, S., \& Beiranvand, N. 2014, MNRAS, 437, 3112
\bibitem[\protect\citeauthoryear{}{}]{} Mosallanezhad, A., Bu, D., \& Yuan, F. 2016, MNRAS, 456, 2877M
\bibitem[\protect\citeauthoryear{}{}]{} Mosallanezhad, A., Khajavi, M., \& Abbassi, S. 2013, RAA, 13, 87M
\bibitem[\protect\citeauthoryear{}{}]{} Narayan, R., Sadowski, A., Penna, R. F., Kulkarni, A. K., 2012, MNRAS, 426, 3241
\bibitem[\protect\citeauthoryear{}{}]{} Narayan, R. 2005, Ap\&SS, 300, 177
\bibitem[\protect\citeauthoryear{}{}]{} Narayan, R., \& Popham, R. 1993, Nature, 362, 820
\bibitem[\protect\citeauthoryear{}{}]{} Narayan, R., \& Yi, I. 1995b, ApJ, 452, 710
\bibitem[\protect\citeauthoryear{}{}]{} Narayan, R., \& Yi, I. 1994, ApJ,428, L13 (NY94)
\bibitem[\protect\citeauthoryear{}{}]{} Narayan, R., \& Yi, I. 1995a, ApJ, 444, 231 (NY95a)
\bibitem[\protect\citeauthoryear{}{}]{} Narayan, R., McClintock J. E., 2008, New Astron. Rev,  51, 733
\bibitem[\protect\citeauthoryear{}{}]{}Novikov I. D., Thorne K. S., 1973, in DeWitt C., DeWitt B., eds, BlackHoles. Gordon \& Breach, New York, p. 345
\bibitem[\protect\citeauthoryear{}{}]{} Pang, B., Pen U. L., Matzner, C. D., Green, S. R., Liebendorfer, M., 2011. MNRAS, 415, 1228
\bibitem[\protect\citeauthoryear{}{}]{} Penna R. F., Sdowski A., Kulkarni A. K., Narayan R., 2013, MN- RAS, 428, Issue 3, 2255- 2274
\bibitem[\protect\citeauthoryear{}{}]{} Pringle, JE. 1981, Annu.Rev. Astron. Astrophys., 19, 137
\bibitem[\protect\citeauthoryear{}{}]{} Rees, M. J., Begelman M. C., Blandford, R. D., Phinney, E. S., 1982, Nature, 295, 17
\bibitem[\protect\citeauthoryear{}{}]{} Sadowski, A., Narayan, R., Penna, R., Zhu, Y. 2013, MNRAS, 436, 3856
\bibitem[\protect\citeauthoryear{}{}]{} Samadi, M., \& Abbassi, S. 2016, MNRAS, 455, 3381S
\bibitem[\protect\citeauthoryear{}{}]{} Samadi, M., Abbassi, S., \& Khajavi, M. 2014, MNRAS, 437, 3124
\bibitem[\protect\citeauthoryear{}{}]{} Samadi, M., Abbassi, S., \& Lovelace, R. V. E. 2017, MNRAS, 470, 2018
\bibitem[\protect\citeauthoryear{}{}]{} Shakura, N. I., Sunyaev, R. A., 1973,  \aap, 24, 337
\bibitem[\protect\citeauthoryear{}{}]{} Stone, J. M., Pringle, J. E., \& Begelman, M. C. 1999, MNRAS, 310, 1002
\bibitem[\protect\citeauthoryear{}{}]{} Tanaka T., Menou K., 2006, ApJ, 649, 345
\bibitem[\protect\citeauthoryear{}{}]{} Xu G., Chen X., 1997, ApJ, 489, L29
\bibitem[\protect\citeauthoryear{}{}]{} Xue L., Wang J.-C., 2005, ApJ, 623, 372
\bibitem[\protect\citeauthoryear{}{}]{} Yuan F., Bu, D. \& Wu, M., 2012, ApJ, 761, 130
\bibitem[\protect\citeauthoryear{}{}]{} Yuan F., Gan Z., Narayan R., Sadowski A., Bu D., Bai X., 2015, ApJ, 804, 101
\bibitem[\protect\citeauthoryear{}{}]{} Yuan F., Wu, M., \& Bu, D. 2012, ApJ, 761, 129
\bibitem[\protect\citeauthoryear{}{}]{} Yuan, F. 2007, in ASP Conf. Ser. 373, The Central Engine of Active Galactic Nuclei, ed. L. C. Ho \& J.-M. Wang (San Francisco, CA: ASP), 95
\bibitem[\protect\citeauthoryear{}{}]{} Yuan, F. 2011, in ASP Conf. Ser. 439, The Galactic Center: A Window to the Nuclear Environment of Disk Galaxies, ed. M. R. Morris, Q. D. Wang, \& F. Yuan (San Francisco, CA: ASP), 346
\bibitem[\protect\citeauthoryear{}{}]{} Yuan, F., \& Bu, D. 2010, MNRAS, 408, 1051
\bibitem[\protect\citeauthoryear{}{}]{} Yuan, F., \& Narayan, R. 2014, ARA\&A, 52, 529
\bibitem[\protect\citeauthoryear{}{}]{} Zeraatgari, F. Z., \& Abbassi, S. 2015, ApJ, 809, 54, (ZA15)
\bibitem[\protect\citeauthoryear{}{}]{} Zhang D., Dai Z. G., 2008, MNRAS, 388, 1409




\end{thebibliography}
\end{document}